\documentclass[letterpaper]{article} 
\usepackage{aaai25-arxiv}  
\usepackage{times}  
\usepackage{helvet}  
\usepackage{courier}  
\usepackage[hyphens]{url}  
\usepackage{graphicx} 
\usepackage{natbib}  
\usepackage{caption} 
\usepackage{array}
\usepackage{amsmath}

\usepackage{algorithm}
\usepackage{algorithmic}

\usepackage{newfloat}
\usepackage{listings}

\usepackage{color}
\newcommand{\todo}[1]{}
\renewcommand{\todo}[1]{{\color{red} TODO: {#1}}}
\DeclareCaptionStyle{ruled}{labelfont=normalfont,labelsep=colon,strut=off} 
\floatstyle{ruled}
\newfloat{listing}{tb}{lst}{}
\floatname{listing}{Listing}
\title{Open-source Polymer Generative Pipeline}
\author {
   Debasish Mohanty\textsuperscript{\rm 1,2},
   Shreyas V\textsuperscript{\rm 1,3},
   Akshaya Palai\textsuperscript{\rm 2},
   Bharath Ramsundar\textsuperscript{\rm 1},
}
\affiliations {
   \textsuperscript{\rm 1}Deep Forest Sciences\\
   \textsuperscript{\rm 2}CIPET:SARP-LARPM\\
   \textsuperscript{\rm 3}BITS Pilani Goa Campus\\
   bharath@deepforestsci.com
}

\begin{document}

\maketitle

\begin{abstract}
Polymers play a crucial role in the development of engineering materials, with applications ranging from mechanical to biomedical fields. However, the limited polymerization processes constrain the variety of organic building blocks that can be experimentally tested. We propose an open-source computational generative pipeline that integrates neural-network-based discriminators, generators, and query-based filtration mechanisms to overcome this limitation and generate hypothetical polymers. The pipeline targets properties, such as ionization potential (IP), by aligning various representational formats to generate hypothetical polymer candidates. The discriminators demonstrate improvements over state-of-the-art models due to optimized architecture, while the generators produce novel polymers tailored to the desired property range. We conducted extensive evaluations to assess the generative performance of the pipeline components, focusing on the polymers' ionization potential (IP). The developed pipeline is integrated into the DeepChem framework, enhancing its accessibility and compatibility for various polymer generation studies.
\end{abstract}

\section{Introduction}

Polymers are macromolecules composed of repeating organic units with diverse applications ranging from manufacturing to drug delivery. Due to the limitations in available organic building blocks with polymerization methods, the range of synthesized polymers is restricted. However, their low cost and ease of production make these macromolecules suitable for various commercial and scientific purposes \citep{polymer_app}. Given the impracticality of experimentally exploring all possible chemical combinations, it is essential to implement artificial intelligence workflows to understand the underlying chemistry and generate polymers with desired properties. Artificial intelligence workflows have led to significant advancements in protein structure analysis \citep{alphafold}, drug discovery \citep{drugreview}, and the generation of hypothetical materials \citep{matgenome}, providing a foundation for their application in polymer chemistry research. Polymers have been structurally featurized to ensure compatibility with neural networks training while preserving maximum chemical detail. Key symbolic representations include Polymer Simplified Molecular Input Line Entry System (PSMILES), BigSMILES \citep{bigsmiles}, fingerprints, molecular graphs, and weighted directed graphs \citep{wdgraph}.


The recent polymer generative methods prioritize output quantity to enhance fingerprint generation, optimize latent spaces for multitask discrimination, and create benchmarking datasets \citep{polybert, 1MRNN}. We propose an open-source pipeline integrating discriminators with generators to produce hypothetical polymers with specified properties. The development of this pipeline involves the following key challenges and approaches:

\begin{itemize}
    \item Discriminators and generators employ various polymer representations, such as monomer SMILES, PSMILES, and molecular graphs. We address this by standardizing these representations and establishing conversion mechanisms between them.
    \item Aligning new target properties with existing generators poses challenges due to compatibility issues between discriminators and generators. We propose training or loading predefined discriminators with custom properties and defining generation rules to maintain pipeline integrity while incorporating new properties.
    \item Existing generative methods generate polymers in large quantities, incurring computational costs proportional to the number of generations. By focusing on polymers with specific properties, we aim to limit the number of generations, reducing computational load and ensuring efficient inference.
\end{itemize}

\section{Related Works}
\subsection{DeepChem and Generative Molecular Models} 
\label{deepchem_gmm} DeepChem~\citep{deepchem} is a versatile open-source Python library tailored for machine learning on molecular and quantum datasets~\citep{deepchem_alt}. Its framework supports applications in drug discovery and biotech~\citep{deepchem_alt}, breaking down scientific tasks into workflows built from core primitives. 
DeepChem has facilitated significant advancements, including large-scale molecular machine learning benchmarks via MoleculeNet~\citep{deepchem_alt}, protein-ligand modeling~\citep{gomes2017atomic}, and generative molecule modeling~\citep{fastflow}. 
The generator models are implemented separately within the DeepChem framework. Additionally, the polymer featurization mechanism, validation workflows, and generative sampler layers have been integrated into DeepChem as part of the development in this work.
\subsection{Reaction-Based Polymer Generation}

In experimental studies, polymers with similar building blocks are often assumed to exhibit similar properties, making the generation of hypothetical polymers from known polymers highly valuable. Conditional polymer classification and polymerization processes have successfully produced valid polymer molecules in several studies. For instance, the SMiPoly library \citep{smipoly} segregates polymerizable functional groups, classifies them, and subjects them to virtual polymerization by specifying starting monomers and reactions. Functional group transformations are then defined, resulting in hypothetical polymer sets. This method follows 19 monomer classes through 22 polymerization rules to form 7 distinct polymer classes. One primary constraint of following this process is the increased likelihood of generating high molecular weight products. A similar approach is employed in the PolyVERSE dataset \citep{polyverse}, where reaction templates, such as ring-opening metathesis polymerization (ROMP), are used to generate high-temperature dielectrics for capacitor materials. The BRICS method simplifies this process by adapting techniques from drug design to generate valid polymer units with minimal manual intervention \citep{polybert}. While BRICS has been effective for small organic molecules and produces valid homopolymers, it falls short in capturing the intricacies of polymerization processes. Despite its ease of use, BRICS does not fully address the complexity required for detailed polymer formation.

\subsection{Neural Network-Based Polymer Generation}

A benchmark PSMILES dataset, PI1M, was developed using an RNN trained on over 12,000 experimental datapoints \citep{1MRNN}. While the RNN successfully captured relationships between PSMILES sequences and generated chemically valid PSMILES, it did not provide insight into the synthetic pathways of the generated polymers. To address this limitation, the open Macromolecular Genome utilized a VAE-based model, Molecule Chef \citep{moleculeChef}, to search for PSMILES with required properties along with the synthetic pathway, and a similar VAE-based model trained on SELFIES. The Molecule Chef architecture classified polymers and indicated their synthetic pathways. The Hierarchical VAE (HierVAE) \citep{hierVAE} approach also proved effective in generating polymer motifs and establishing bonds sequentially. Although HierVAE  was primarily designed to identify functional motifs of macromolecules, it has also been applied to polymer generation with SMILES.

\section{Methodology}

\begin{figure*}[!h]
\centering
\includegraphics[width=0.9\textwidth]{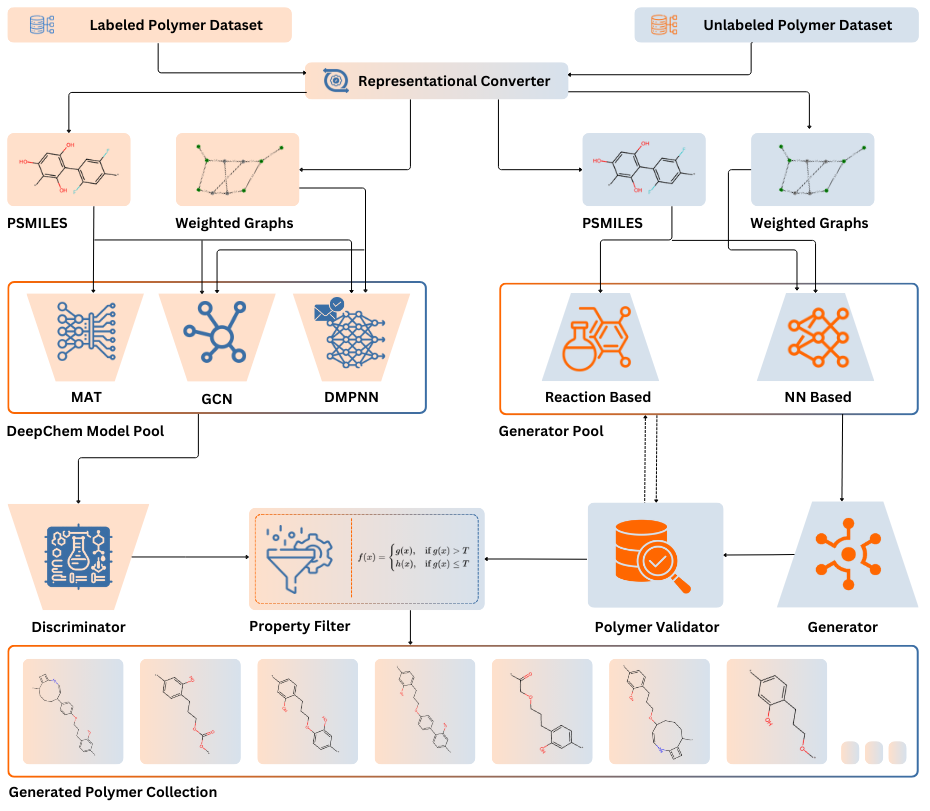}
 \caption{Overview of the polymer generative pipeline to generate polymer representation of PSMILES and weighted directed graphs utilizing a DeepChem model pool (comprising of Molecule Attention Transformer - MAT \cite{MAT}, Graph Convolution Network  - GCN \cite{gcn}, Directed Message Passing Neural Network - DMPNN \cite{DMPNN}), and generator pool (comprising of Reaction Based, Neural Network Based Architecture). Consecutively, it is passed through a polymer validator and filter to narrow down the generation to targeted candidates.
}
\label{fig1}
\end{figure*}

This section outlines the methodologies employed across different stages of the pipeline. The pipeline integrates representational variations with conversion mechanisms, paired with discriminator and generator architectures, to generate hypothetical polymers. A filtration process is then applied to refine the generated polymers to those most relevant. An overview of these components is provided in Figure \ref{fig1}. The key methodologies driving these components are detailed below.

\begin{figure*}[t]
    \centering
    
    \textbf{A} $$[indexed PSMILES_1].[indexed PSMILES_2]|[Monomer Fraction_1]|[Monomer Fraction_2]|
$$
$$<[Bond Index_1]-[Bond Index_2]:[W_{1-2}]:[W_{2-1}] ... <[Bond Index_n] - [Bond Index_m]:[W_{n-m}]:[W_{m-n}]  $$ 
\textbf{B}
$$[indexed PSMILES_1].[indexed PSMILES_1]|0.5|0.5|
$$
$$<[Bond Index_1]-[Bond Index_2]:[W_{1-2}]:[W_{2-1}] ... <[Bond Index_n] - [Bond Index_m]:[W_{n-m}]:[W_{m-n}]  $$ 
\caption{\textbf{A.} The syntax of the weighted directed graph representation. \textbf{B.} The syntax WDG converted from PSMILES of homopolymers}

\label{wdformat}
\end{figure*}
\subsection{Representational Variations and Conversions}

\begin{figure*}[!ht]
\centering
\includegraphics[width=0.9\textwidth]{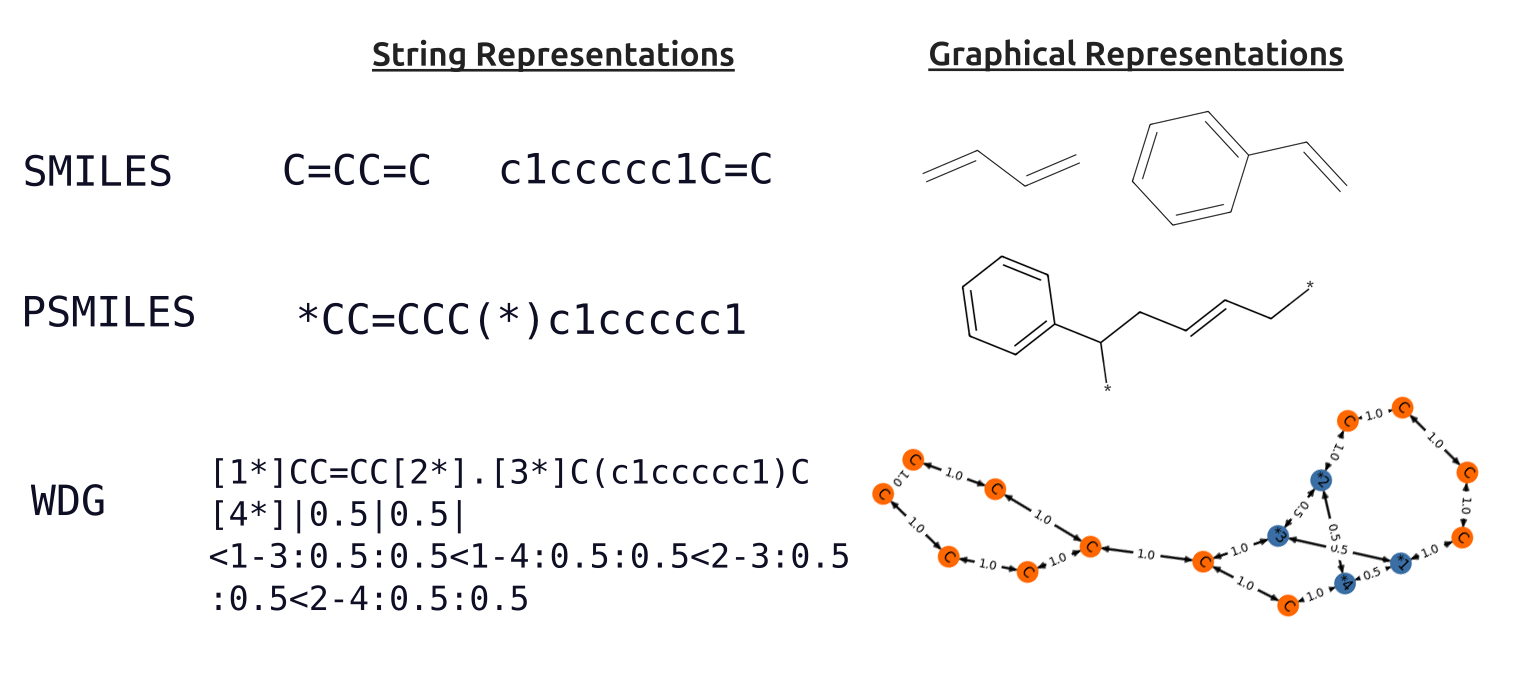}
\caption{The sample representational variations of the Buna-S (rubber) polymer and its monomers.}
\label{repr_var}
\end{figure*}

The sequential representation of polymers can be expressed using monomer SMILES, PSMILES, and BigSMILES. PSMILES is used as a core representation in the pipeline for standardization and dataset compatibility. In homopolymer representations, each repeating unit is analogous to monomer SMILES, with "*" indicating open bonds. In the case of copolymers, the CRU is represented similarly with the ``*" character for open bonds. The weighted graph is stored as a sequential string with specific notations, similar to PSMILES. However, instead of ``*", open bonds are indexed as $[*: i]$ or $[i*]$, where $i$ is an integer, allowing weights to be assigned to relevant bonds. This representation aligns with graph-based neural networks and specifically weighted directed message-passing neural networks (DMPNN) \citep{graphrepr}. The complete weighted graph representation consists of three parts: (1) a string representing the constituent atoms, (2) fractions of repeating units separated by "$|$" (e.g., "$|0.5|0.5|$"), and (3) polymer rules, indicated by "$<$", mapping open bond indices to weights that correspond to polymer stoichiometry variations. An abstract of the weighted directed graph string syntax is illustrated in Figure \ref{wdformat}. To provide a detailed understanding, we simulated the formation of Buna-S using the SMILES representations of the monomers 1,3-butadiene and styrene. The corresponding representational variations, including PSMILES and weighted directed graphs (WDG), are illustrated in figure \ref{repr_var}.

A conversion mechanism was necessary to align data with relevant architectures. We generated the constitutional repeating unit (CRU) by converting weighted directed graphs into PSMILES. This involved converting the molecule into an editable form (via RDKit), joining relevant bonds to form the CRU from monomer SMILES, and removing indexed SMARTS notations to create valid PSMILES. Monomer fractions and weights were retained as metadata, which could later regenerate the original Weighted Directed Graphs. For raw PSMILES, the monomer was repeated twice to emulate connections between repeating units, open bonds were indexed, and stoichiometric relationships (random, alternating, block) were used to calculate and assign weights at the tail of the representation, as shown in Figure \ref{wdformat}.

\subsection{Reaction-Based Generative Method}

Among the various reaction-based generative methods, BRICS was chosen for our application due to its ability to incorporate the inherent chemistry of monomer molecules, independent of the polymerization process. Additionally, the computational load of PSMILES can be adjusted based on the complexity of the input molecule. BRICS is inherently incompatible with PSMILES due to the wildcard "*" notation, which leads to chemically invalid bond breaking and recomposing. To resolve this, we first convert PSMILES into valid Molecular SMILES by replacing the "*" with a virtual atom, represented as "[At]". BRICS then processes the modified SMILES to detect substructures and generate valid SMILES efficiently. After generating the SMILES, we validate the polymer representation by checking the count and position of virtual atoms. Upon validation, the virtual atoms are replaced with the wildcard notation. This approach allows BRICS to be effectively utilized to generate hypothetical polymers.

\subsection{Neural Network-Based Generative Method}

The input format for both PSMILES and Weighted Directed Graphs (WDG) is sequential, stored and retrieved as strings. Based on this, we trained a sequential neural network to capture representational patterns in PSMILES and WDG, aiming to generate plausible polymer structures. Variational Autoencoders (VAE) and Generative Adversarial Networks (GAN) are commonly employed for generating synthetic data across domains, including image and molecular generation. In our approach, we focused on sequential relevance, evaluating the impact of the same neural architecture on both representations. We used a BERT tokenizer to convert the string-based PSMILES and WDG formats into tokens. The model architecture includes an embedding layer, followed by Long Short-Term Memory (LSTM) blocks, and a fully connected layer that outputs sequences matching the BERT tokenizer's vocabulary. The LSTM mechanisms—input gate (Equations \ref{eq:inputLSTM1}, \ref{eq:inputLSTM2}, \ref{eq:inputLSTM3}), forget gate (Equation \ref{eq:forgetLSTM}), and output gate (Equation \ref{eq:outputLSTM})—are governed by the following equations:

Input Gate:
\begin{equation}
    f_i = \sigma(W_i \cdot [h_{t-1},x_t] + b_i)
    \label{eq:inputLSTM1}
\end{equation}

\begin{equation}
    \hat{C} = \tanh(W_c.[h_{t-1}, x_t] + b_c)
    \label{eq:inputLSTM2}
\end{equation}

\begin{equation}
    C_t = f_t \odot C_{t-1} + i_t \odot \hat{C}_t
    \label{eq:inputLSTM3}
\end{equation}

Forget Gate:
\begin{equation}
    f_t = \sigma(W_f \cdot [h_{t-1},x_t] + b_f)
    \label{eq:forgetLSTM}
\end{equation}

Output Gate:
\begin{equation}
    o_t = \sigma(W_o \cdot [h_{t-1},x_t] + b_o)
    \label{eq:outputLSTM}
\end{equation}

Here, $W$ represents the weight matrix for each gate, where i, f, and o correspond to the input, forget, and output gates, respectively. $h_{t-1}$ denotes the hidden state from the previous time step, and $x_t$ is the current input. $\hat{C}$ refers to the cell state vector derived from the tanh activation in the previous iteration. The symbol $\odot$ indicates element-wise multiplication, while $\sigma$ represents the sigmoid activation function used in the corresponding equations.

The mechanism of LSTM effectively propagates the significance of previous tokens in a sequence, making it well-suited for capturing the importance of wildcards in both PSMILES and weighted directed graph (WDG) representations \citep{LSTM}. However, in longer sequences, the forget gate can reduce the retention of information from distant tokens, limiting its effectiveness for extended sequence lengths. Despite this, LSTMs are computationally more efficient than transformers, especially when dealing with larger latent spaces. Given the relatively short sequence lengths for both PSMILES and WDGs, LSTMs were a good starting point. 

The model employs BERT case-aware tokenization with a vocabulary size of 28,996. A 128-dimensional embedding layer is followed by two stacked unidirectional LSTM layers, each with a hidden dimension of 256. A fully connected neural network then maps the LSTM output to the tokenizer’s vocabulary size. Sigmoid and tanh activation functions are used exclusively within the LSTM layers (as shown in equation \ref{eq:inputLSTM1}, \ref{eq:inputLSTM2}, \ref{eq:forgetLSTM}, \ref{eq:outputLSTM}). The architecture is trained on approximately 1 million PSMILES and 42,000 WDG string representations.

\subsection{Generation Validation}

Several benchmarking protocols have been developed for validating molecular generation systems, with SMILES serving as the standard input format for such evaluations \citep{moses, guacamol}. Due to the partial misalignment of SMILES with PSMILES and WDG formats, we implemented a custom benchmarking framework to assess the validity, uniqueness, and novelty of the generative model. For PSMILES validation, we used RDKit's molecular structure parser, while for WDG, a string validator, and a sample featurizer were employed to ensure the generation of chemically valid graphs. The uniqueness score, expressed as a percentage, was determined by calculating the proportion of unique molecules among the valid generations. The novelty was assessed by comparing the unique molecules to the training data, ensuring that the generator produced new molecules distinct from the training set. Two validation batches, one with 1,000 generations and another with 10,000, were performed. The 1,000-generation batch tested the validity of the model, while the 10,000-generation batch examined the system's ability to maintain uniqueness over extended iterations. These three metrics—validity, uniqueness, and novelty—were employed to evaluate the performance of the generative model in this study.

\subsection{Property Discriminative Methods}

\begin{figure}[t]
\centering
\includegraphics[width=0.8\columnwidth]{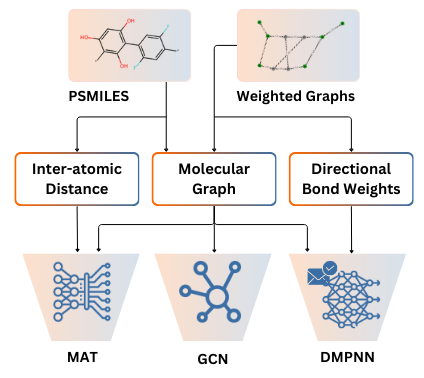}
\caption{Representational variation and the discriminator types of Molecule Attention Transformer (MAT), Graph Convolution Network (GCN), and Directed Message Passing Neural Network (DMPNN) with their corresponding intermediary input forms}
\label{discrim_dist}
\end{figure}

Property prediction tasks based on PSMILES and WDG representations can leverage discriminative neural networks. PSMILES provides sufficient information to create a molecular graph, while WDG encodes bond occurrence probabilities. Three state-of-the-art architectures were considered for polymer discrimination: Molecule Attention Transformer (MAT) \citep{MAT}, Graph Convolution Network (GCN) \citep{gcn}, and Directed Message Passing Neural Network (D-MPNN) \citep{DMPNN}. MAT employs transformer attention mechanisms along with inter-atomic distances and molecular graph structures for property prediction, though inter-atomic distance conflicts with wildcards limit minor PSMILES data use. GCN applies convolutional operations to graph nodes, allowing atom-level (node) analysis of target properties. Both PSMILES and WDG can be represented as graphs, aligning well with GCN. D-MPNN passes auxiliary information as messages through the graph, capturing additional factors beyond node location. This architecture is suited for PSMILES and WDG due to the inclusion of wildcard locations and bond weights as message data. The schema of the representational methods and corresponding data form mapped with the discriminator architecture is shown in Figure \ref{discrim_dist}.

\subsection{Targeted Polymer Filtration with Discriminator}

\begin{figure}[t]
\centering
\includegraphics[width=0.8\columnwidth]{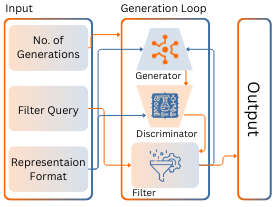}
\caption{The pipeline cycle of generating polymers from input elements to generation loop applied with the filter mechanism.}
\label{pipeline_loop}
\end{figure}

A filtering mechanism was implemented to generate a specified number of polymers based on a discriminator filter through iterative steps. A loop is established over the generator-discriminator stack, where hypothetical polymers are generated and evaluated by the discriminator to assess target properties. The filter is represented as a logical string containing comparison operators ($<$, $>$) and floats to define conditions. The generated polymer's properties are checked against the filter to determine if they meet the desired criteria. The loop terminates once the target polymer count is reached, with a timeout to prevent indefinite execution in cases of narrow filter criteria. The mechanism has been visualized in Figure \ref{pipeline_loop}.

\section{Experiments}
\subsection{Datasets}

The application required datasets that were both experimentally derived and representationally convertible for the discriminator, as it focuses on predicting experimental target values. For the generator, datasets comprising chemically valid polymer representations are essential.
Auxiliary details are added or omitted during the conversion between PSMILES and WDG. To address this, we selected the WDG dataset as the base for our experiments, converting it to PSMILES for implementation. Due to the addition of synthetic data when converting raw PSMILES into WDG, we prioritized WDG as the base representation and reduced the data to derive corresponding PSMILES. Given the limited availability of raw WDG datasets due to their niche usage, we used a dataset of 42,966 copolymers (42K dataset) collected from the same study that introduced weighted directed graphs for molecular ensembles \citep{wdgraph}. For the generator, we employed the BRICS method, which is primarily suited for PSMILES. As BRICS can fragment smaller molecules into valid components, we used experimental data from the converted 42K WDG dataset to train the generator and compare its performance with neural networks. Like most neural network architectures, the generative pipeline requires a substantial amount of data for producing valid outputs. Therefore, we utilized a synthetic dataset of 1 million PSMILES data points generated via BRICS for PolyBERT model development. PolyBERT's source data contains 100 million data points, but we used a subset to train our model and validate generation using PSMILES. For training the WDG generative model, we relied on the 42K dataset. 


\subsection{Baseline}
The study using the same WDG dataset serves as a basis for comparison in our work. The original study employed the Weighted Directed Message Passing Neural Network (w-DMPNN) for WDG representation but utilized the standard DMPNN with monomer representations for analysis \citep{wdgraph}. Besides DMPNN, Random Forest and Neural Networks using molecular fingerprints were also explored, with DMPNN being the top-performing model. The base study evaluated performance using root mean squared error (RMSE) and R-squared ($R^2$) metrics. The baseline DMPNN with monomer representation achieved an RMSE of 0.156 and an $R^2$ of 0.89. However, their analysis provided limited numerical metrics for polymer generation models. Polymer generation benchmarking, such as in the case of HierVAE, was conducted using SMILES input and output, but the metrics are not directly comparable due to representational conflicts \citep{hierVAE}. Therefore, we designed a custom benchmarking setup tailored to PSMILES and WDG to evaluate generation models based on uniqueness, novelty, and validation metrics across different methods and input types.

\subsection{Training and Evaluation}
Training was conducted on MAT, GCN, and DMPNN architectures for the development of the discriminator, with an 80:20 train-test split. Molecular featurization for MAT, GCN, and DMPNN was performed using DeepChem’s Molecular Featurizers. For WDG, a specialized graph representation was required to incorporate atomic details and auxiliary data. To address this, a custom graph featurizer was implemented to generate modified graphs compatible with both GCN and DMPNN. Following featurization, the models were trained for 50 epochs, using mean squared error (MSE) as the loss function. Additional metrics were used to assess performance, including the R-squared ($R^2$) value and mean absolute percentage error (MAPE).

The BRICS method, as a reaction-based generative approach, does not require training but instead relies on a sufficient number of input PSMILES. We utilized 6110 unique PSMILES, derived from the 42K dataset after excluding stoichiometry and chain variations, to apply the BRICS generation pipeline. BRICS generates chemically valid molecules; however, only those with exactly two wildcards are eligible for polymer chain formation, requiring a filter to isolate potential monomers. After generation, the valid PSMILES were evaluated for uniqueness and novelty.

For neural network-based generation, BERT tokens were used for both PSMILES and WDG. The LSTM model was trained for 5 epochs using 1 million PSMILES and for 50 epochs using 42K WDG data. Generation was performed in batches of 1000 (run 10 times) and 10000 (run once). The generated outputs were evaluated separately for validity, uniqueness, and novelty, with inference carried out using a filtering mechanism.

\subsection{Validator and Filter Alignment}

The validation process is applied in two stages: during generator evaluation and in the pre-filtering generation steps. The process begins with a specified target for polymer candidate generation, where each iteration of the generation loop passes through a validator. After validation, a filtration logic based on a string query is applied. The validator and filter stack continuously update the generated candidates' count, halting once the desired number is reached. The validator for polymers was directly implemented from DeepChem utilization classes. The filter layer is modified to assess the system's complexity and time efficiency, particularly when the filter represents a narrow window for the target property or when the target value falls slightly outside the discriminator's training range.

\subsection{Pipeline Performance Assessment}

The pipeline's performance was evaluated based on the time required to generate a specified number of samples relative to the filter window for the target variable. Initially, the filter was set narrowly around the mean of the ionization potential data, with two margin values, 0.1 and 0.01, representing the range above and below the mean. The experiment was conducted by iteratively expanding the filter window, increasing the margin from 1x to 5x in five steps. A margin of 0.1 was used to ensure the pipeline addressed a broad range of discriminator targets, while 0.01 enforced a narrow filter. 

\section{Experimental Results}
\subsection{Discriminator Evaluation}

\begin{figure}[!ht]
\centering
\includegraphics[width=0.9\columnwidth]{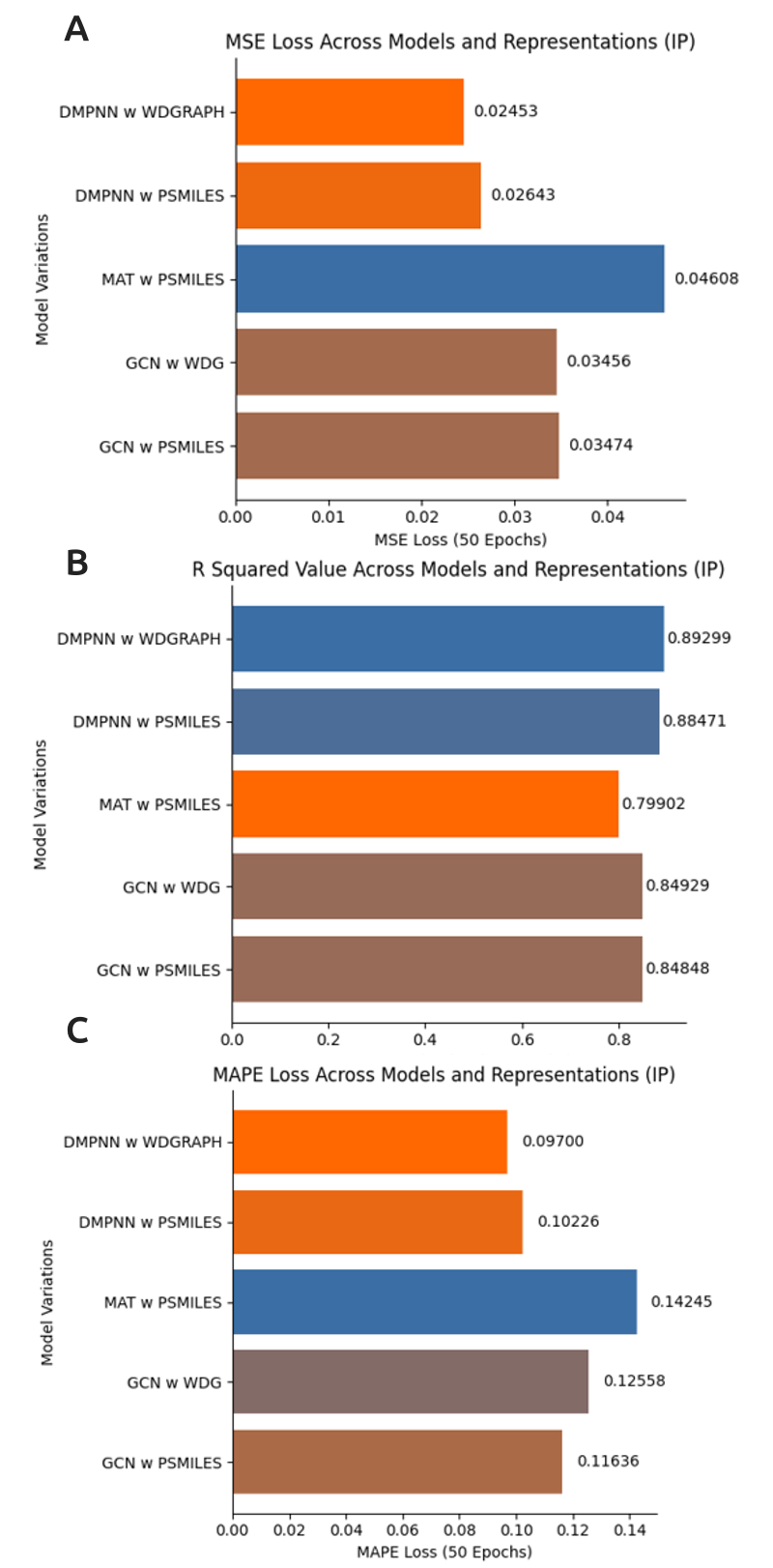}
\caption{The bar plot between PSMILES and Weighted Directed Graph representation mapped with discriminative
architecture Directed Message Passing Neural Network (DMPNN) and Graph Convolution Network (GCN) comprising
Mean Squared Error (MSE) loss (section A), R Squared Value (section B), and Mean Absolute Percentage
Error (MAPE) (Section C)
}
\label{lossfig}

\end{figure}

By aligning representational variations with discriminator architectures, five discriminators were developed: three using PSMILES with GCN \citep{gcn}, MAT \citep{MAT}, and DMPNN \citep{DMPNN}, and two using WDG with GCN and DMPNN. The MAT model, which partially aligns with PSMILES, had the highest RMSE (0.214) and a lower $R^2$ score (0.80) with the ionization potential (IP) dataset. The GCN model demonstrated similar RMSE values for both PSMILES and WDG, with a slight RMSE reduction for WDG. Corresponding $R^2$ scores were 0.84 and 0.85 for PSMILES and WDG, respectively. DMPNN, designed for WDG, performed better with an RMSE of 0.156 and $R^2$ of 0.89 compared to 0.162 and 0.88 for PSMILES. The baseline DMPNN with monomer representation in WDG had an RMSE of 0.16, slightly outperforming the PSMILES DMPNN. Including additional weight data in the DMPNN featurization using DeepChem reduced the RMSE to 0.156 and improved the $R^2$ score to 0.89. The evaluation metrics from the test set are summarized in Table \ref{table-disc-compare}. Mean Absolute Percentage Error (MAPE) and $R^2$ score are calculated as additional metrics. The metrics values for each discriminator variation has been visualized in Figure \ref{lossfig}.

\begin{table}[h!]
\centering
\begin{tabular}{l|c|c}
\textbf{Model Variation} & \textbf{RMSE} & \textbf{$R^2$ Score} \\
\hline
GCN with PSMILES & 0.186 & 0.84 \\
GCN with WDG & 0.185 & 0.85 \\
MAT with PSMILES & 0.214 & 0.80 \\
DMPNN with PSMILES & 0.162 & 0.88 \\
DMPNN with WDG & \textbf{0.156} & \textbf{0.89} \\
Baseline DMPNN with WDG & 0.16 & 0.88 \\
\hline
\end{tabular}
\caption{The calculated RMSE and $R^2$ Score for different models with different representations compared with the baseline DMPNN implemented with monomer representation using weighted directed graphs dataset (42K) targeting Ionization Potential (IP)}
\label{table-disc-compare}
\end{table}


\begin{table}[!h]
\centering
\begin{tabular}{l|c|c|c}
\noalign{\vskip 0.3em} 
& Valid & Unique & Novel \\ [0.3em]
\hline
\noalign{\vskip 0.3em} 
\textbf{Generator Models} & \multicolumn{3}{|c}{1K Gen} \\
\hline
LSTM w WDG & 67\% & 79\% & 7\%  \\
LSTM w PSMILES & 51\% & 100\% & 100\%  \\
BRICS w PSMILES & 100\% & 100\% & 100\% \\
\hline
\noalign{\vskip 0.3em} 
& \multicolumn{3}{|c}{10K Gen} \\
\hline
LSTM w WDG & 67.5\% & 37\% & 16.5\% \\
LSTM w PSMILES & 50.7\% & 100\% & 99.9\% \\
\hline
\end{tabular}
\caption{Metrics of Validity, Uniqueness, and Novelty for Generator Models (LSTM with WDG, LSTM with PSMILES, and BRICS with PSMILES) averaged over 10 sample runs of 1,000 and 10,000 generations.}
\label{table-disc-compare}
\end{table}

\subsection{Generator Evaluation}

\begin{figure}[!h]
\centering
\includegraphics[width=0.9\columnwidth]{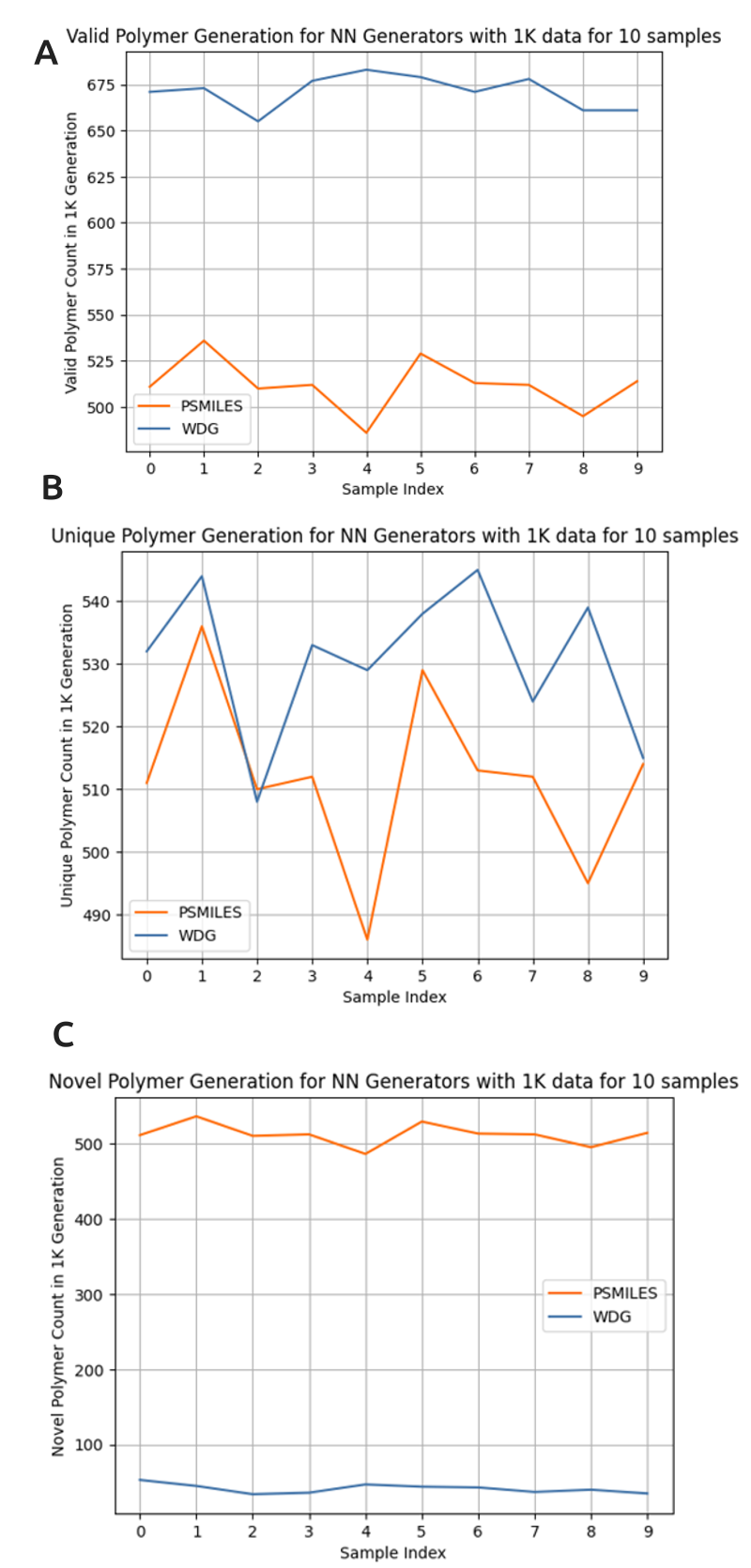}
\caption{Line plot comparison of generator performance with PSMILES and WDG representations, showing (A) the number of valid polymers, (B) the number of unique polymers, and (C) the number of novel polymers over 1,000 generations across 10 sample runs. 
}
\label{generator_comp}

\end{figure}

The generator was evaluated over 1,000 (1K) generations using both LSTM and BRICS models across 10 samples. The valid, unique, and novel polymer generation metrics for each sample and varied LSTM representations are shown in Figure \ref{generator_comp}. On average, 671 valid polymers were generated in 1K generations with the Weighted Directed Graph (WDG) representation, of which 531 were unique and 41 were novel, corresponding to valid, unique, and novel rates of 67\%, 79\%, and 7\%, respectively. In a separate experiment with 10,000 (10K) generations using WDG, 6,757 valid polymers were produced, including 2,496 unique and 412 novel polymers. The results indicate that, while valid generations remain consistent at larger scales, the number of unique polymers decreases, yet novel polymers remain stable.

With LSTM and PSMILES, 1K generations yielded 512 valid molecules, all of which were unique and novel, resulting in a 51\% valid score and 100\% unique and novel scores. At 10K generations, valid scores remained at 50.7\% (5,071 valid molecules), with 100\% uniqueness and a slight reduction in novelty to 99.9\% (5,068 novel molecules). These results highlight significant differences between LSTM with WDG and LSTM with PSMILES. The smaller latent space from LSTM with WDG (trained on 42K samples) compared to LSTM with PSMILES (trained on 1M samples) leads to repeated candidates, lowering uniqueness and novelty. BRICS, a reaction-based generator, consistently produced 100\% valid, unique, and novel polymers over 1K generations. Testing with 10 PSMILES inputs resulted in 1,476 valid PSMILES, of which the first 1,000 were used for evaluation, all being unique and novel. 

\subsection{Time Constraint Analysis}

The filter margins for polymer candidate generation were set at 0.1 and 0.01. In both cases, the LSTM with PSMILES required less time per cycle due to more efficient generation and shorter PSMILES sequences. With a 0.1 filter margin, the WDG pipeline took 45.7 seconds to generate 10 candidates, while with a 0.01 margin, it took 506 seconds. In contrast, the PSMILES pipeline required 11.1 seconds with a 0.1 margin and 93 seconds with a 0.01 margin. Regardless of representation, narrowing the filter margin from 0.1 to 0.01 increased the time required by approximately 10-fold.

\begin{figure*}[]
\centering
\includegraphics[width=0.9\textwidth]{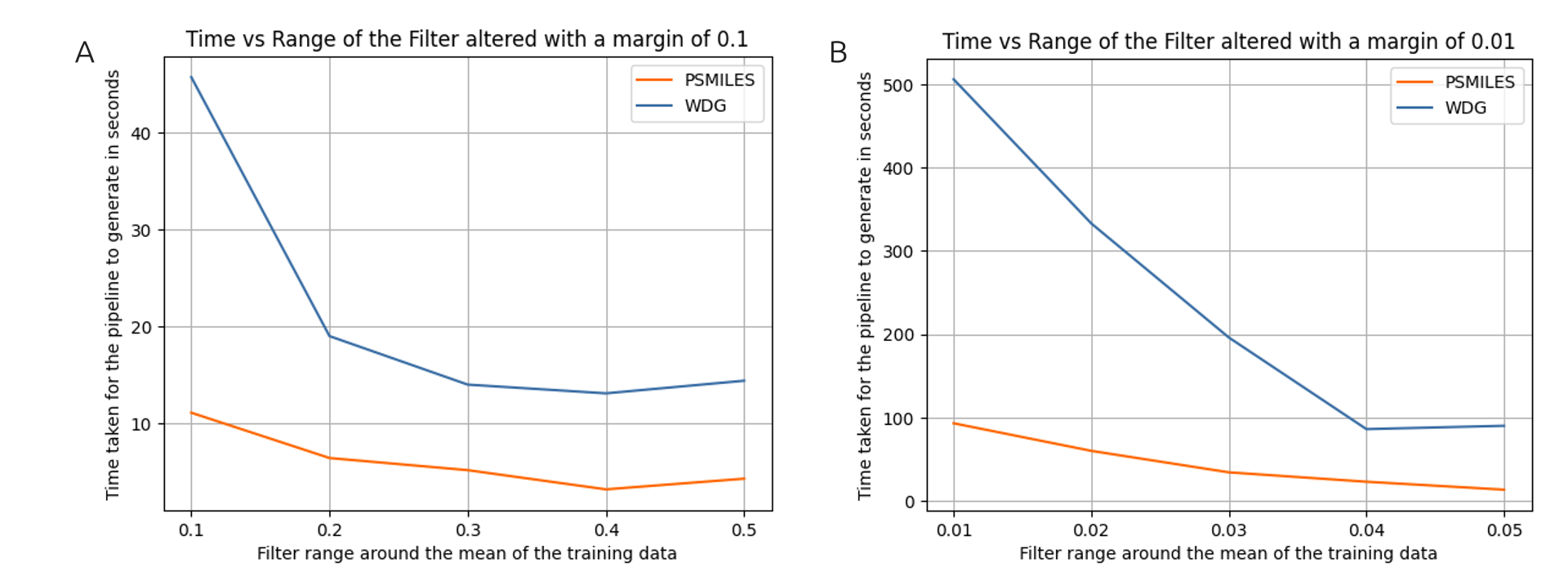}
\caption{ Line plot showing the time taken for the pipeline to generate 10 polymers with a filter tolerance of 0.1 (A) and 0.01 (B), across 5 intervals from the target property mean. The plots compare the time taken across pipelines with different representational variations.
}
\label{time_comp}

\end{figure*}

Time constraints appeared to stabilize after the 4th iteration for both representations. For the 0.1 margin, WDG took 14.4 seconds and PSMILES took 4.3 seconds in the 5th iteration, compared to 13.1 and 3.2 seconds in the 4th iteration, respectively. With a 0.01 margin, WDG took 86 seconds and PSMILES took 22.8 seconds in the 4th iteration, followed by 90 seconds and 13.3 seconds in the 5th iteration. While time reduction stabilized after the 4th iteration for WDG with both margins, the PSMILES pipeline continued to show gradual time reduction beyond the 4th iteration in the 0.01 margin case. The time reduction trends are visualized in Figure \ref{time_comp}.

\section{Conclusion}

This research presents a pipeline that integrates open-source discriminative and generative model architectures capable of iteratively generating hypothetical polymers for a specific target property. The pipeline accommodates various polymer representation formats, enabling both neural network-based and reaction-based generation approaches. Experiments were conducted to evaluate the performance of the generators, discriminators, and filter mechanisms in producing relevant polymer candidates within a defined timeframe. Ionization potential (IP) was used as the target property for training and implementation. The pipeline components demonstrated state-of-the-art performance, successfully predicting polymer candidates for the specified property. Performance variations due to representational differences were analyzed and addressed.


\bibliography{aaai25, ref}

\end{document}